\documentclass[submission,copyright,creativecommons]{eptcs}
\providecommand{\event}{PLACES 2016}

\usepackage{alltt}
\usepackage{amsmath}
\usepackage{amssymb}
\usepackage{pygmentex}
\usepackage{float}
\usepackage{microtype}

\newcommand{\PL}{{\sc Syndicate}}

\title{From Events to Reactions: A Progress Report}

\author{Tony Garnock-Jones
  \institute{Northeastern University, Boston, Massachusetts, USA}
  \email{tonyg@ccs.neu.edu}}

\def\titlerunning{From Events to Reactions: A Progress Report}
\def\authorrunning{T. Garnock-Jones}

\setpygmented{texcomments,lang=racket,font=\ttfamily\footnotesize,escapeinside=||,hidealllines=true}
\newcommand{\SOURCEFIG}[1]{\inputpygmented[topline=true,bottomline=true]{#1}\vspace{-0.75em}}
\newcommand{\SOURCE}[1]{\inputpygmented[]{#1}\vspace{-0.4em}}

\begin{document}
\maketitle


\begin{abstract}
 \PL{} is a new coordinated, concurrent programming language. It
 occupies a novel point on the spectrum between the shared-everything
 paradigm of threads and the shared-nothing approach of actors.
 \PL{} actors exchange messages and share common knowledge via a
 carefully controlled database that clearly scopes conversations. This
 approach clearly simplifies coordination of concurrent activities.
 Experience in programming with \PL{}, however, suggests a need to
 raise the level of linguistic abstraction. In addition to writing event handlers
 and managing event subscriptions directly, the language will have to
 support a reactive style of programming. This paper presents
 event-oriented \PL{} programming and then describes a preliminary
 design for augmenting it with new reactive programming constructs.
\end{abstract}


\section{Introduction}


Interactive programs must keep track of their conversations with
entities in the outside world, demultiplexing incoming events
and maintaining internal state associated with each
concurrent conversation. For example, web servers manage multiple
conversations with browsers that come and go unpredictably.
Even sequential programming languages face the challenges of this form of concurrency.

\PL{}~\cite{GarnockJones2016,GarnockJones2014} aims
to ease management of a program's interactions with the
outside world.
The language is centred around functional actors which exchange events and
actions with their context.
The Actor model~\cite{Hewitt1973} gives each actor a piece of
private state, eschews shared state, and offers point-to-point
delivery of messages to peers in a single, flat group.
\PL{} loosens these restrictions with three novel features: {\em
  multicast} message delivery instead of point-to-point
messaging;\footnote{More properly, multicast delivery {\em subsumes}
  point-to-point messaging: the latter can be encoded using the
  former.} a carefully controlled {\em shared dataspace} allowing
peers to collaboratively maintain shared state; and {\em hierarchical
  containment} allowing decomposition of one actor into many, able to
communicate internally via a private message bus and group-private
shared dataspace.

An actor in \PL{} is either a {\em leaf actor}, with behaviour given
by a programmer-supplied function, or a {\em network actor}, with
behaviour specified by the language itself. A network actor contains
and manages other actors, providing shared message bus and dataspace
services to them.
Messages are routed to actors based on pattern-matching {\em subscriptions} rather than actor IDs.

The active subscriptions in a network are just one kind of shared
state. In the actor model, the (implicit) routing table connecting
peers is the {\em only} shared state.\footnote{While the {\em model}
  enforces isolation, practical actor {\em languages} almost always
  include ad-hoc mechanisms for sharing state. Often this
  is not an explicit design decision but a reflection of the
  impracticality of removing or controlling existing shared-state
  facilities in an underlying language such as Java, Javascript or C++.} \PL{}
generalises the routing table and allows actors to place arbitrary
data, known as {\em assertions}, within it. The resulting structure is
dubbed a {\em dataspace}. Subscriptions---assertions of {\em interest}
in data of a particular shape---serve not only to route messages
toward an actor, but to match assertions in the dataspace. Networks
notify interested contained actors of the existence of assertions
relevant to their interests, sending {\em state change notification}
events as assertions come and go and as actors assert new and retract
old subscriptions. In this way, dataspaces are similar to Linda's
tuplespaces~\cite{Gelernter1985}. However, unlike the ``generative''
tuples of Linda, assertions in \PL{} do not have independent
lifetimes. Instead, an assertion remains in a network's dataspace only
so long as some contained actor continues to assert it. When actors
terminate, cleanly or abnormally, all their assertions are removed,
and interested parties are notified of the change. This yields a
failure signalling mechanism that is part of \PL{}'s approach to
lifecycle management of components.

Networks deliver events to their actors, which reply with actions to
be interpreted by the network. Messages are sent and new actors
created roughly as in the traditional actor model, but assertions,
including subscriptions, are managed by {\em state change
  notifications}. An actor sends such notifications to its containing
network to manipulate the set of active assertions associated with
the actor and shared with the group. Symmetrically, networks send
such notifications to contained actors to keep them up-to-date as
assertions relevant to their interests come and go. State change
notifications take the form of {\em patches} against a previous
assertion set, describing assertions to be added to and removed from
the set.

\PL{} programmers write behaviour functions that react to incoming
events and yield updated actor state and a sequence of actions to
perform. However, directly programming to this basic
interface quickly becomes inconvenient. To fully achieve its goal of easing
the management of concurrent interactions between a program and its environment,
\PL{} needs higher-level constructs.

\section{\PL{} By Example}
\label{sec:syndicate-by-example}

\begin{figure}[t]
  \SOURCEFIG{SOURCE-prospect-core.rkt}
  \caption{Selected \PL{} constructs as expressed in Racket}
  \label{fig:core-syntax}
\end{figure}



The primary implementation of \PL{} is an embedded language based on
Racket~\cite{plt-tr1}. The language provides data structures and
syntactic constructs corresponding to \PL{}'s mathematical
definition~\cite{GarnockJones2016}. Some of these are shown in
Figure~\ref{fig:core-syntax}.
Events and actions are represented by Racket {\tt struct}s, and
behaviour functions associated with leaf actors are represented by
Racket functions.

Programming in \PL{} involves the design of {\em protocols} specifying
message types, interaction patterns, and roles that protocol
participants must play. In addition, unlike protocols for actor
systems, \PL{} protocols must specify data to be placed in the shared
dataspace along with instructions for its maintenance. Each network
actor thus embodies a protocol instance, and its contained actors are
the protocol participants.


\begin{figure}[t]
  \SOURCEFIG{SOURCE-bank-account-plain.rkt}
  \caption{Bank account actor behaviour functions}
  \label{fig:bank-account-behaviours}
\end{figure}

As is traditional,
our
example is a drastically oversimplified simulated bank
account. The protocol involves three roles: account manager, account
updater, and account observer. Only one account manager is permitted
per protocol instance, but any number of updaters and observers may
exist. The account manager maintains an {\tt account} structure in the
shared dataspace to describe the account balance. Account observers
monitor this structure, tracking balance changes.
Account updaters send {\tt deposit} messages, which are
interpreted by the account manager. The structures concerned are
defined as follows:
\SOURCE{SOURCE-bank-account-protocol.rkt}

The role of account manager is performed by the leaf actor created by
interpretation of the following {\em spawn} action specifying a
behaviour function, an initial state, and a collection of startup
actions:
\SOURCE{SOURCE-bank-account-manager-plain.rkt}


The new actor's state is the account balance, initially zero. Its sole
startup action is a state change notification---a patch---asserting
two facts: that the current balance (in an {\tt account} structure) is
zero, and that the actor is interested in receiving {\tt
  deposit} messages. Its behaviour function ({\tt manager}, from Figure~\ref{fig:bank-account-behaviours})
lets it react to
such messages: after computing an updated balance, it yields its new
state (namely, the new balance) along with a patch action updating
the contents of the shared dataspace with the new value. The patch is
computed as the cumulative effect of two simpler patches: the first
retracts the old assertion, and the second asserts the new.

An observer that prints the new balance each time it changes is
created as follows:
\SOURCE{SOURCE-bank-account-observer-plain.rkt}

The observer is locally stateless:
its state throughout its lifetime is {\tt (void)}.
Its initial action
asserts its interest in assertions of shape {\tt (account ?)}. In
response, its containing network---the implicit {\em ground network}
at the outermost layer of the running actor hierarchy---sends
state change notification events keeping it informed of the {\tt
  account} assertions in the dataspace. The observer responds to each
such event by {\em projecting out} and printing the {\tt balance}
field of each newly-added {\tt account} assertion. To do this, it
uses a projection specification
including a {\em capture} operator, written {\tt (?!)}. The
specification and the set of newly-added assertions are given to the
function {\tt project-assertions}. The result is a
Racket set containing the captured values, if any. The observer loops
over the set's contents, printing each value.

Finally, we exercise the account via
an actor performing the role of an account updater:
\SOURCE{SOURCE-bank-account-updater-plain.rkt}

It would be disastrous to send {\tt deposit} messages before
the account manager was listening for them. The updater therefore
delays sending until interest in the messages to be sent is detected
in the shared dataspace. Its initial patch, then, asserts interest in
{\em assertions of interest in {\tt deposit} messages}. Once
its behaviour function receives a patch containing a non-empty added
set, it can conclude that some peer is ready to receive {\tt
  deposit} messages. It then sends two such messages before
terminating cleanly.

The result of all this is that the account's balance, reflected as
{\tt account} assertions in the shared dataspace, starts at zero,
climbs to 100, and finally drops to 70. No further activity occurs.








\section{Reactive Constructs for \PL{}}


The example, simple though it is, highlights several \PL{} idioms
deserving of linguistic support. The preliminary ``reactive'' constructs
presented in this section aim to capture the essence of such idioms:

\begin{description}

\item[Input.] Actors wish to receive messages, but must manage
    assertions of interest in messages separately from the code
    responding to each message.

    Similarly, actors wish to track the
    contents of the shared dataspace, and must not only manage
    assertions of interest in the relevant portion of the dataspace
    but also apply projections to the sets of assertions they are
    informed about.


\item[Output.] Actors not only send messages, but maintain sets of
    assertions relevant to peers.

\item[Demultiplexing.] Actors engage in multiple conversations at
    once, each handling one facet of the actor's overall interface to
    the outside world. Occasionally, the assertions of one ongoing
    conversation will {\em overlap} with those of another. The actor
    must take care not to inadvertently affect other conversations
    when adjusting one conversation's assertion set. Also, the actor
    must carefully examine each incoming event to determine which of
    its active conversations it might relate to.

\item[Auxiliary conversations.] Events can trigger complex chains of
    actions. For example, as part of reacting to an event, an actor
    may wish to call some other actor in
    the system, sending a message and continuing where it left off
    upon receiving a reply. It must then keep track of waiting
    continuations.

\end{description}

Figure~\ref{fig:actor-syntax} shows the proposed additions to the
language.
In
the functional interface to \PL{}, actors manipulate {\em action
  structures} describing intended actions as values. By contrast, the
new {\tt send!} form is imperative, sending a message as a side
effect. Likewise, {\tt actor} creates, as a side effect, a new actor
that executes the contained expressions sequentially.

At the heart of the new design is the {\tt state} form and the related concept
of {\em facets}. A {\tt state} form is a {\em blocking} construct that
stays active until one of its {\em termination events} occurs. When
this happens, the expressions associated with the particular
termination event are evaluated and the resulting values returned to
the waiting continuation of the {\tt state}. The {\tt until} and {\tt
  forever} forms are special cases of {\tt state} with one and zero
termination events, respectively.

While a {\tt state} is running, its associated facets are active. A
facet is either an {\tt assert}, which keeps an assertion set in
the shared dataspace up-to-date as its actor's state evolves, or an
{\tt on}, which responds to incoming events. If
\pyginline|#:collect| is present in a {\tt state} form, the bindings
next to it are in scope in all of the {\tt state}'s facets. Each {\tt
  on} facet's body must return as many values as there are
\pyginline|#:collect|ed bindings, making the entire {\tt state} form
into a kind of fold over events.

There are four kinds of events that can trigger evaluation of an {\tt
  on} facet's body or one of the termination conditions of a {\tt
  state}. A {\tt message} event entails a subscription to the
associated message pattern, and is triggered when a matching message
comes in. An {\tt asserted} event is fired for {\em each} assertion
added to the shared dataspace and matching the given pattern. A {\tt
  retracted} event is similar, but for assertions removed from the
dataspace. Patterns in these event specifications use ``{\tt \$}'' to
introduce binders, and ``{\tt \_}'' as a wildcard. Finally, {\tt
  rising-edge} events trigger whenever, after some event arrives from
the outside world, the given Racket expression evaluates to a true
value, having previously evaluated to false.

\begin{figure}[t]
  \SOURCEFIG{SOURCE-actor-core.rkt}
  \caption{``Reactive'' \PL{} constructs. $\left<\cdots\right>$ indicates optional syntax.}
  \label{fig:actor-syntax}
\end{figure}

The following example demonstrates the basic ideas:
\begin{pygmented}[]
  (actor (send! 'starting)
         (define final-count
           (state [#:collect [(count 0)]                               ;; (a,b)
                   (assert (list 'incrs-seen-so-far count))            ;; (b)
                   (on (message 'incr) (+ count 1))]                   ;; (a)
             [(rising-edge (>= count 5)) (send! 'too-many)    count]   ;; (c)
             [(message 'interrupt)       (send! 'interrupted) count])) ;; (c)
         (send! 'finished))
\end{pygmented}
\vspace{-0.5em}
The actor shown sends a message, \pyginline|'starting|, and then enters a {\tt
  state} which (a) counts \pyginline|'incr| messages, (b) maintains an
assertion in the shared dataspace describing the count seen so far,
and (c) waits either for the count to reach $5$ or for an
\pyginline|'interrupt| message. Once one of the latter events happens,
a message is sent before the final count is returned to the
continuation of the {\tt state}. Once the state terminates, the
\pyginline|'incrs-seen-so-far| assertion is retracted along with the
event handlers monitoring \pyginline|'incr| messages and the
termination conditions. The actor then terminates after sending a
\pyginline|'finished| message.

Together, the constructs of Figure~\ref{fig:actor-syntax} support the
idioms listed in the introduction to this section. Each {\tt on} facet is automatically
translated into code for maintaining appropriate assertions of
interest, depending on the kind of event acting as its trigger. Facets
awaiting assertions and retractions automatically project incoming
patches and iterate over the resulting sets. Each {\tt assert}
facet automatically reevaluates its associated expression and, when it changes, updates
the shared dataspace. Run-time support routines
multiplex the assertions of the active facets, ensuring that they do
not interfere, and dispatch incoming events to relevant facets.
Finally, the blocking nature of {\tt state} expressions lets event
handler code suspend until some condition obtains. Unlike Erlang's
selective receive construct, the actor's active facets remain
responsive while a sequence of expressions is suspended waiting for
termination of a {\tt state}. In other words, ongoing conversations
are not disrupted by the temporary existence of an auxiliary
conversation. An analogous distinction can be seen between modal and
non-modal dialogs in graphical user interfaces.

\paragraph*{Revisiting the bank account example.}

The new constructs shorten our example and clarify the purpose of the
code. The account manager is expressed as follows:
\SOURCE{SOURCE-bank-account-manager-actor.rkt}

The account observer becomes:
\SOURCE{SOURCE-bank-account-observer-actor.rkt}

Finally, our updater is simply:
\SOURCE{SOURCE-bank-account-updater-actor.rkt}

The {\tt until} expression blocks waiting for an
\pyginline|(observe (deposit _))| assertion to
appear in the shared dataspace.
The \pyginline|(on (message (deposit $amount)) ...)| facet in
the account manager automatically establishes just such an assertion,
thereby indirectly releasing the balance-alteration messages, now
known to have a recipient.




\paragraph*{Implementation.}

The prototype implementation is a library of macros and
support routines that may be used to write any given leaf actor. The
support routines use Racket's delimited continuations~\cite{Flatt2007}
to provide a direct-style, imperative veneer atop the functional
interface between a leaf actor and its containing network demanded by
\PL{}.


\section{Related Work}

\PL{}'s dataspaces must be discussed in terms of comparison not only
to other {\em models} of concurrent computation but also to practical
actor {\em languages}. At the model level, dataspaces are related to
Linda's {\em tuplespaces}~\cite{Gelernter1985}, and its state change
notifications to Erlang's failure-signalling mechanisms, {\em links}
and {\em monitors}~\cite{Armstrong2003}. Tuples in Linda's tuplespaces
are ``generative''; that is, once created, they take on independent
existence. \PL{}'s assertions, however, survive only as long as some
actor continues to assert them. The resulting failure-signalling facility
is comparable to Erlang's links and monitors, which
allow actors to subscribe to messages signalling the termination of
specified peers. Cast in \PL{} terms, if an actor $X$ asserts its
existence, {\tt (ok $X$)}, then monitoring peers can react to {\tt
  (retracted (ok $X$))}.

At the language level, dataspaces are quite different to forms of
general-purpose shared state seen in actor languages. \PL{} actors can
only affect a dataspace by placing assertions into it. In exchange for
this severe restriction, automatic retraction of assertions on actor
termination protects the structure from damage.
By contrast, Erlang's ETS tables are effectively read/write hash
tables shared among processes. They suffer many of the same problems
of other kinds of shared, mutable storage, including no reliable way
to detect or repair damage after abnormal exit of a process. Languages
such as Akka, working within systems making liberal use of mutable,
shared state, suggest as a matter of convention the use of immutable
data structures, but cannot enforce this. The resulting shared state
facility is orthogonal to the language's actor model. It is worth
noting that our Racket and Javascript \PL{} implementations also go
beyond the \PL{} model: they suggest, but do not enforce, restrictions
on use of mutable state. Finally, mention must be made of the language
Pony, which employs {\em deny capabilities}~\cite{Clebsch2015} to
efficiently and safely share access to objects. While \PL{} lacks
shared memory of this kind, it could be used as a substrate for Pony's
distributed resource management protocols, or Pony could be used to
implement a single \PL{} leaf actor as a group of collaborating Pony
actors.

Turning to the reactive constructs described in this paper, Hancock's
language ``Flogo II''~\cite{Hancock2003} embodies ideas that are in
some ways similar. Hancock introduces a ``step/process distinction''
and aims for a ``unified para\-digm'' midway between ``declarative''
and ``procedural'' approaches to interactive programming. ``Steps'' in
Flogo II are analogous to imperative actions in Racket, and
``processes'' are roughly analogous to facets, \PL{}'s ongoing
reactive constructs within blocking {\tt state} expressions. \PL{} is
clearly aiming at similar territory to Flogo II; however, the
treatment of private and shared state in the Flogo II prototype is
quite different, and the language lacks a full treatment of
concurrency, message passing and error recovery. The reactive
constructs presented here, by contrast, rest upon \PL{}'s
event-oriented formal semantics.

The term ``reactive'' is used in many contexts. It is used here to
connote similarity to languages such as Dedalus~\cite{Alvaro2009} and
to models such as functional reactive
programming~\cite{Elliott1997,Cooper2006} and even spreadsheets.
Dedalus programs are structured around a distributed Datalog variant
similar to \PL{}'s dataspace. They react to changes in the Datalog
database in some ways like \PL{} actors reacting to changes in the
shared dataspace. Functional reactive programs construct dataflow
graphs atop streams of time-varying values; if we view the
time-varying set of assertions matching some expressed interest as a
representation of such a stream, we can imagine a \PL{} actor
implementing the computation at a dataflow graph node and publishing
its result downstream as an assertion in turn. An entire program would
then be a collection of such actors. Finally, the spreadsheet model,
in which cells subscribe to updates from their peers and publish their
own contents in turn, can be directly encoded in \PL{}. It remains
future work to make these connections to other reactive designs
precise.

\section{Conclusion}

This paper has given a brief overview of event-oriented programming in
\PL{}, and motivated and introduced a preliminary design for a
``reactive'' style of programming in \PL{}. The design captures
several important idioms of \PL{} programming, and makes programs
employing these idioms both shorter and easier to read and understand.

The presented design is not only preliminary, but partial: ongoing
work includes the exploration of constructs for automating the
integration of incoming state change notifications to yield various
aggregate structures and to compute predicates over the dataspace. In
addition, the {\tt state} form always creates a new actor, even when
used in tail position. A future refinement should lift this
limitation. Finally, future work will also include formal
specification of the mapping from the proposed constructs to \PL{}'s
core semantics.

\paragraph*{Acknowledgements.}
This work was supported in part by several NSF grants. The author
would like to thank the anonymous reviewers for their feedback,
comments and suggestions. In addition, many thanks to Matthias
Felleisen, Sam Caldwell, and the participants of NU PLT’s coffee
round.

\paragraph*{Resources.}
The programs shown in this paper, along with other examples and
resources, may be found via
\url{www.ccs.neu.edu/home/tonyg/places2016/}.


{
\footnotesize
\nocite{*}
\bibliographystyle{eptcs}
\bibliography{places2016}
}


\appendix

\section{Extended example: ``File System''}
\label{appendix:fs}

In this section, we examine a ``file system'' actor based on a sketch
given in the original \PL{} paper~\cite{GarnockJones2016}. The responsibility
of the ``file system'' is to store the contents of a named collection
of {\em files}, and to provide some way of reading, writing and
deleting such files.

The protocol accepted by the file system is simple:
\SOURCE{SOURCE-fs-protocol.rkt}

There is no special message or assertion type needed to read files.
Interested actors simply subscribe to {\tt file} structures with the
appropriate name and are notified as the file content changes.

\begin{figure}[t]
  \SOURCEFIG{SOURCE-fs-hll.rkt}
  \caption{File System example: ``Reactive'' \PL{} Implementation}
  \label{fig:fs-hll}
\end{figure}


The complete reactive file system implementation is shown in Figure~\ref{fig:fs-hll}.
Let us examine its behaviour when run alongside an actor that
monitors the contents of {\tt novel.txt}:
\begin{pygmented}[]
  (actor (forever (on (asserted (file "novel.txt" $text))
                      (display "The novel's content is: ") (write text) (newline))))
\end{pygmented}

When this novel-monitoring actor starts, it asserts its interest in
the content of the novel, which causes an assertion {\tt (observe
  (file "novel.txt" \_))} to be placed in the dataspace. This in turn
triggers the file system's {\tt (on (asserted (observe (file \$name \_))) ...)} facet, which
enters a state in which it asserts the current contents of the
novel---initially \pyginline|#f|, signifying a nonexistent file---and
watches for updates.


If now some actor executes
\pyginline|(send! (save (file "novel.txt" "It was a dark and stormy night")))|,
then {\em both} the file system actor itself {\em and} the
{\tt until} clause maintaining the {\tt file} assertion will receive a
message. The file system actor updates its private state---a hash
table mapping file names to file contents---and the {\tt until}
clause updates its own state, which causes its \pyginline|(assert (file name content))|
facet to be re-evaluated. The novel-monitor in turn receives the
update as a new assertion notification, and the change is displayed on
the console for the user to see.

The novel-reader shown is simple-minded. A more sophisticated actor
might become bored with its reading and retract its interest in the
novel; the {\tt until} clause in the file system would then
terminate, retracting the assertion of the novel's contents in turn.
The file-system actor thus not only provides a key-value store and a
change-notification service, but also manages a kind of {\em cache},
reflecting certain values from the authoritative store (its {\tt
  files} state variable) into the shared dataspace. The lifetime of
each activation of the {\tt until} clause scopes the lifetime of the
corresponding cache entry and of the ongoing conversation, simple
though it is in this case, with the monitoring actor.

An approximately equivalent ``{\em non}-reactive'' program is shown in
Figure~\ref{fig:fs-lll}. The file-system actor has behaviour function
{\tt file-system-event-handler}, while each activation of the inner
{\tt until} clause is modelled as a separate actor with behaviour {\tt
  file-observation-event-handler}. Missing is the linkage between the
two classes of actor that is automatically supplied in the
``reactive'' case. Interesting points of difference include the manual
repetition of patterns (e.g.\ lines 5 and 20, 13 and 21, and 15 and 22),
managed automatically by the reactive constructs; the manual update of
assertions in the shared space (lines 26--27), again managed
automatically by the ``assert'' facet in the reactive implementation;
the explicit use of projection and iteration over assertion sets
(lines 5 and 32), managed automatically by the reactive {\tt until}
form; and, not least, the great difference in size. The
``non-reactive'' implementation is 38 lines long, while the reactive
implementation is only 9 lines long.

\begin{figure}
  \SOURCEFIG{SOURCE-fs-lll.rkt}
  \caption{File System example: ``Non-reactive'' \PL{} Implementation}
  \label{fig:fs-lll}
\end{figure}


\end{document}